\documentclass[twocolumn,aps,showpacs,prb,tightenlines,amsmath,amssymb]{revtex4}
\usepackage{graphicx}
\usepackage{amssymb}
\usepackage{dcolumn}
\usepackage{amsmath}
\usepackage{pifont}
\usepackage{bm}
\usepackage{colordvi}
\newcommand{\bgreek}[1]{\mbox{\boldmath$#1$\unboldmath}}

\begin{document}

\title{Topological superconductor with a large Chern number and a large bulk excitation gap in single layer graphene}
\author{L. Wang}
\thanks{wlf@mail.ustc.edu.cn}
\affiliation{Hefei National Laboratory for Physical Sciences at
Microscale and Department of Physics,
University of Science and Technology of China, Hefei,
Anhui, 230026, China}
\author{M. W. Wu}
\thanks{mwwu@ustc.edu.cn}
\affiliation{Hefei National Laboratory for Physical Sciences at
Microscale and Department of Physics, University of Science and
Technology of China, Hefei, Anhui, 230026, China}

\date{\today}

\begin{abstract}
We show that a two-dimensional topological superconductor (TSC) can be realized
in a hybrid system with a conventional $s$-wave superconductor proximity-coupled
to a quantum anomalous Hall (QAH) state from the Rashba and exchange effects in
single layer graphene. With very low or even zero doping near the Dirac
  points, i.e., two inequivalent valleys, this TSC has a Chern number as large
  as four, which supports four Majorana edge modes. More importantly, we show 
that this TSC has a robust topologically nontrivial bulk excitation
    gap, which can be larger or even one order of magnitude larger than the proximity-induced
superconducting gap. This unique property paves a way for the application
  of QAH insulators as seed materials to realize robust TSCs and Majorana
  modes.
\end{abstract}

\pacs{73.43.-f, 81.05.ue, 71.10.Pm, 74.45.+c}

\maketitle
\section{INTRODUCTION}
Majorana modes can naturally exist in topological
superconductors (TSCs).\cite{kitaev,alicea1, beenakker,hasan,qi} The
intrinsic TSC has been predicted to exist in
superconducting Sr$_2$RuO$_4$ with $p$-wave paring state.\cite{ivanov,mackenzie} 
However, this has not yet been experimentally confirmed. Recently, many 
efforts have been devoted to design artificial TSCs.\cite{fu,sau,sau2,alicea2,lutchyn,alicea3,halperin,stanescu,mourik,yzhou,bysun,nadj,hui,
dumitrescu,jianli,qi2,jwang,poyhonen,jjhe,tewari,tewari2,mdiez,wong,haim,rontynen,dutreix,jianli2}
So far, most studies focus
on the effective $p$-wave superconductors in hybrid systems with conventional
$s$-wave superconductors in proximity to strong topological insulators,\cite{fu}
semiconductors with strong spin-orbit coupling (SOC),
\cite{sau,sau2,alicea2,lutchyn,alicea3,halperin,stanescu,mourik,yzhou,bysun,jjhe,tewari,tewari2,mdiez,wong,haim,dutreix}
or ferromagnetic atom
chains.\cite{nadj,hui,dumitrescu,jianli,poyhonen,rontynen,jianli2} 
Some attention has also been paid to the conventional $s$-wave superconductors coupled to quantum anomalous Hall
(QAH) insulators such as topological insulators with magnetic
dopants.\cite{qi2,jwang} 
Among all the above TSCs, multiple spatially overlapping
  Majorana modes, which greatly benefit the transport properties, can only coexist in one-dimensional (two-dimensional) TSCs
  belonging to Class BDI \cite{poyhonen,haim,dumitrescu,hui,jianli,jjhe,tewari,tewari2,mdiez,wong} 
(D \cite{qi2,jwang,rontynen,jianli2}) with integer topological
invariant.\cite{schnyder} In reality, the one-dimensional TSCs in Class BDI can easily reduce to
  the ones indexed by Class D with zero or one Majorana
  mode.\cite{poyhonen,haim,dumitrescu,hui,tewari,tewari2,mdiez,wong} As for the two-dimensional TSCs in Class D, the number of the
  Majorana modes or the Chern number is limited upto two.\cite{qi2,jwang,jianli2} More Majorana modes or larger Chern numbers are
  limited by large chemical potential (i.e., very high doping) and an overall much smaller bulk excitation gap than the proximity-induced
  superconducting gap.\cite{rontynen,jianli2}

In this work, we show that a two-dimensional TSC
can be realized in a hybrid system with a conventional $s$-wave superconductor
proximity-coupled to a QAH state \cite{qiao} due to the Rashba SOC \cite{Rashba} and exchange field in
single layer graphene. Interestingly, with very low or even zero doping near the Dirac points,
i.e., two inequivalent valleys, the TSC from the QAH state has a Chern number
reaching as large as four, hosting four Majorana edge modes. More importantly, these Majorana modes are
protected by a bulk excitation gap, which can be larger or even one order of
magnitude larger than the superconducting gap from the proximity
effect. This is in strong contrast to the case of effective $p$-wave
  superconductors where the excitation gap is always smaller than the
  superconducting gap.\cite{sau,sau2,alicea2,lutchyn,alicea3,halperin,stanescu,mourik,yzhou,bysun}
  As the large topologically nontrivial gap has been shown to be probably most important for applications in topological
  insulators,\cite{hasan,qi} topological crystalline insulators,\cite{cniu} and QAH
  insulators,\cite{cniu,qiao2,gangxu} our finding, i.e., reporting a large bulk
  excitation gap in the TSC is crucial to the field of TSCs and Majorana
  modes. This paves a way to obtain robust TSCs and Majorana modes using the QAH
  states. We also address the experimental feasibility of the TSC
from the QAH state.

This paper is organized as follows. In Sec.~II, we present our model and lay out
the tight-binding Hamiltonian of single layer graphene. Then, we calculate the
topological invariant in Sec.~III. We further presents the results on the phase
diagram, Majorana edge states and bulk excitation gap in Sec.~IV. Finally, we summarize and discuss in Sec.~V.

\section{MODEL AND HAMILTONIAN}
The real-space tight-binding Hamiltonian of single layer graphene
with the Rashba SOC, exchange field and proximity-induced $s$-wave superconductivity is given by \cite{qiao,kane,ezawa}
\begin{eqnarray}
H&=&-t\sum_{\langle i,j\rangle\alpha}c^{\dagger}_{i\alpha}c_{j\alpha}
+i\lambda\sum_{\langle i,j\rangle\alpha\beta}{({\bgreek
  \sigma}^{\alpha\beta}\times {\bf d}_{ij})}_zc^{\dagger}_{i\alpha}c_{j\beta}\nonumber\\
&&\mbox{}-\mu\sum_{i\alpha}c^{\dagger}_{i\alpha}c_{i\alpha}
+V_z\sum_{i\alpha}c^{\dagger}_{i\alpha}\sigma_z^{\alpha\alpha}c_{i\alpha}\nonumber\\
&&\mbox{}+\Delta\sum_{i}(c^{\dagger}_{i\uparrow}c^{\dagger}_{i\downarrow}+{\rm H.c.}),\label{hamil}  
\end{eqnarray}
where $\langle i,j\rangle$ represents the nearest-neighboring sites and 
$c_{i\alpha}\ (c^{\dagger}_{i\alpha})$ annihilates (creates) an electron with
spin $\alpha$ at site $i$. The first
term stands for the nearest-neighbor hopping with $t=2.7\ $eV \cite{neto} being the hopping
energy. The second term denotes the Rashba SOC with $\lambda$, ${\bgreek \sigma}$ and
${\bf d}_{ij}$ representing the coupling strength, Pauli matrices for real spins and a unit vector
from site $j$ to site $i$, respectively. $\mu$ in the third term is the chemical
potential. $V_z$ ($\Delta$) in the fourth (fifth) term corresponds to 
exchange field (superconducting gap from the proximity effect). 

To start, we transform the Hamiltonian of Eq.~(\ref{hamil}) to the
Bogoliubov-de Gennes (BdG) one in the momentum space. Specifically, 
\begin{eqnarray}
H=\frac{1}{2}\sum_{\bf k}\Phi^{\dagger}_{\bf k}H_{\rm BdG}({\bf k})\Phi_{\bf k}
\end{eqnarray}
where $\Phi^{\dagger}_{\bf k}=(\psi^{\dagger}_{{\rm A}\uparrow}(\bf
k),\ \psi^{\dagger}_{{\rm B}\uparrow}(\bf k),\ \psi^{\dagger}_{{\rm A}\downarrow}(\bf
k),\ \psi^{\dagger}_{{\rm B}\downarrow}(\bf k),\ \psi_{{\rm A}\downarrow}(-{\bf
  k}),\\ \psi_{{\rm B}\downarrow}(-{\bf k}),\ -\psi_{{\rm A}\uparrow}(-{\bf
  k}),\ -\psi_{{\rm B}\uparrow}(-{\bf k}))$ with $\psi^{\dagger}_{i\alpha}({\bf k})$
creating an electron with spin $\alpha$ and momentum ${\bf k}$ counted from the
momentum $\Gamma$ at sublattice $i$ ($i={\rm A}$, ${\rm B}$) and 
\begin{eqnarray}
H_{\rm BdG}({\bf k})=\left(\begin{array}{cc}
H_e({\bf k})-\mu & \Delta\\
\Delta & \mu-\sigma_yH_e^*(-{\bf k})\sigma_y\label{BdG}
\end{array}\right).
\end{eqnarray}
$H_e({\bf k})$ represents tight-binding Hamiltonian without the $s$-wave
superconductivity, which can be written as
\begin{eqnarray}
H_{e}({\bf k})=\left(\begin{array}{cccc}
V_z & f({\bf k}) & 0 & h_1({\bf k})\\
f^*({\bf k}) & V_z & h_2^*({\bf k}) & 0\\
0 & h_2({\bf k}) & -V_z & f({\bf k})\\
h_1^*({\bf k}) & 0 & f^*({\bf k}) & -V_z\label{hamilnosc}
\end{array}\right)
\end{eqnarray}
where $f({\bf k})=-t[(2\cos \frac{k_x}{2}\cos \frac{k_y}{2\sqrt{3}}+\cos
\frac{k_y}{\sqrt{3}})-i(2\cos \frac{k_x}{2}\sin \frac{k_y}{2\sqrt{3}}-\sin
\frac{k_y}{\sqrt{3}})]$, $h_1({\bf k})=-\lambda [(\cos \frac{k_x}{2}+\sqrt{3}\sin
\frac{k_x}{2})\sin \frac{k_y}{2\sqrt{3}}+\sin \frac{k_y}{\sqrt{3}}]-i\lambda [-\cos \frac{k_y}{\sqrt{3}}+\cos
\frac{k_y}{2\sqrt{3}}(\cos \frac{k_x}{2}+\sqrt{3}\sin \frac{k_x}{2})]$ and
$h_2({\bf k})=\lambda [(\sqrt{3}\sin \frac{k_x}{2}-\cos \frac{k_x}{2})\sin
\frac{k_y}{2\sqrt{3}}-\sin \frac{k_y}{\sqrt{3}}]+i\lambda [\cos \frac{k_y}{\sqrt{3}}-\cos
\frac{k_y}{2\sqrt{3}}(\cos \frac{k_x}{2}-\sqrt{3}\sin \frac{k_x}{2})]$. Note that the lattice
constant is set to be unity in the calculation for simplicity. 

\section{TOPOLOGICAL INVARIANT}
Before investigating the topological properties of
$H_{\rm BdG}({\bf k})$, we first identify the gap closing conditions.  The gap closing of the BdG Hamiltonian $H_{\rm BdG}({\bf k})$ is equivalent to
the existence of bulk zero energy states due to particle-hole symmetry. The
condition for bulk zero energy states is obtained by calculating ${\rm
  det}(H_{\rm BdG})=0$. We find that
the gap closes at the momenta $\Gamma$ (single one), $M$ (three
inequivalent ones) and $K$ (two inequivalent ones) points with the corresponding conditions given by $(\mu\pm
3t)^2=V_z^2-\Delta^2$, $(\mu\pm t)^2=V_z^2-\Delta^2$ and $\mu^2=V_z^2-\Delta^2$,
respectively. It is noted that $+$ ($-$) stands for lower (higher) energy band at
the momentum $\Gamma$ or $M$. The detailed calculation is shown in Appendix~\ref{appB}. Obviously, our system is topologically trivial 
in the case of $|V_z|<|\Delta|$. As for $|V_z|\ge |\Delta|$, we have ten critical
chemical potentials in order, i.e., $\mu_{1,2}=3t\pm
\sqrt{V_z^2-\Delta^2}$, $\mu_{3,4}=t\pm \sqrt{V_z^2-\Delta^2}$, 
$\mu_{5,6}=\pm \sqrt{V_z^2-\Delta^2}$, $\mu_{7,8}=-t\pm \sqrt{V_z^2-\Delta^2}$,
and $\mu_{9,10}=-3t\pm \sqrt{V_z^2-\Delta^2}$ by assuming $|V_z|,|\Delta|\ll t$,
which divide the system into eleven topological regimes.   

These topological regimes are characterized by the Chern number $C_1$ 
since $H_{\rm BdG}(\bf k)$ belongs to Class D with integer topological
invariant.\cite{schnyder} $C_1$ can be calculated by \cite{ghosh}
\begin{eqnarray}
C_1=\frac{1}{2\pi}\int_{\rm BZ}d^2{\bf k}f_{xy}({\bf k})
\end{eqnarray} 
with the Berry curvature
\begin{eqnarray}
f_{xy}({\bf k})=i\sum_{m,n}(f_m-f_n){u_m^{\dagger}({\bf
    k})[\partial_{k_x}H_{\rm BdG}({\bf k})]u_n({\bf k})}\nonumber\\
\mbox{}\times {u_n^{\dagger}({\bf k})[\partial_{k_y}H_{\rm BdG}({\bf k})]u_m({\bf k})}/{[E_m({\bf k})-E_n({\bf k})]^2}.
\end{eqnarray}
Here, $u_m({\bf k})$ is the $m$-th eigenvector of $H_{\rm BdG}({\bf k})$ with the
corresponding eigenvalue being $E_m({\bf k})$; $f_m=1\ (0)$ for occupied
(empty) band. The Chern number of all topological regimes is given by
\begin{eqnarray}
C_1=\left\{
     \begin{array}{ll}
      1\ (-1), &\mu_2<\mu<\mu_1\\
      -3\ (3), &\mu_4<\mu<\mu_3\\
      4\ (-4), &\mu_6<\mu<\mu_5\\
      -3\ (3), &\mu_8<\mu<\mu_7\\
      1\ (-1), &\mu_{10}<\mu<\mu_9\\
      0, &\mbox{other regimes}
    \end{array}
    \right.
\end{eqnarray}
when $V_z>0\ (V_z<0)$. It is seen that $|C_1|=1\ (3)$ near the momentum $\Gamma\
(M)$ point, which is consistent with the number of the zero energy states in
Ref.~\onlinecite{dutreix}. These Majorana modes require very large chemical
potential (of the order of eV), clearly unachievable experimentally. It is noted
that the study on the Majorana modes near the Dirac points is absent in
Ref.~\onlinecite{dutreix}. In this work, with very low or even zero doping near
the Dirac points, i.e., $K$ (two inequivalent ones), 
we have a Chern number as large as four. 

\section{Results}
\subsection{Phase diagram}
\begin{figure}[bth]
\centering
\includegraphics[width=6.cm]{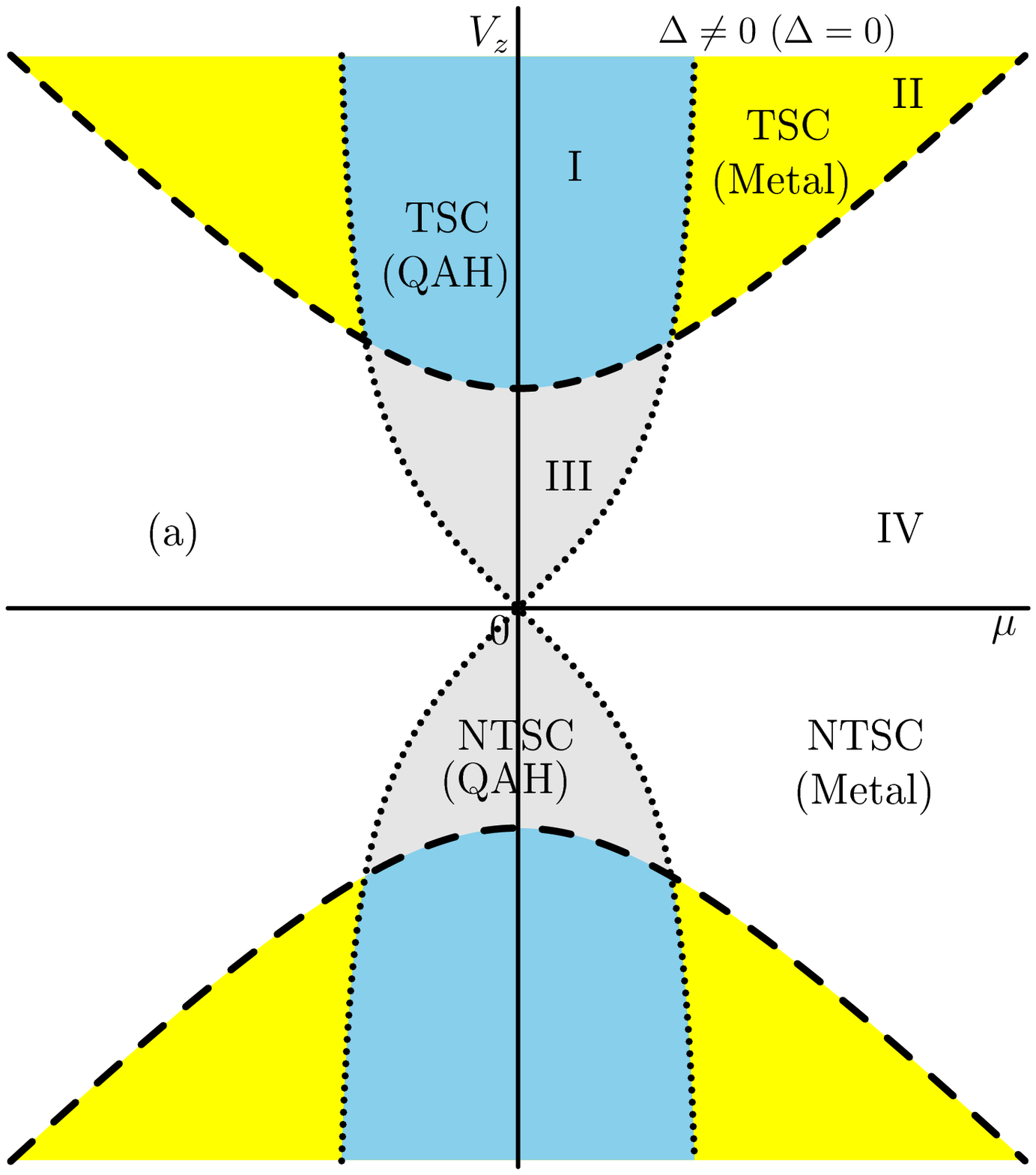}
\includegraphics[width=8.5cm]{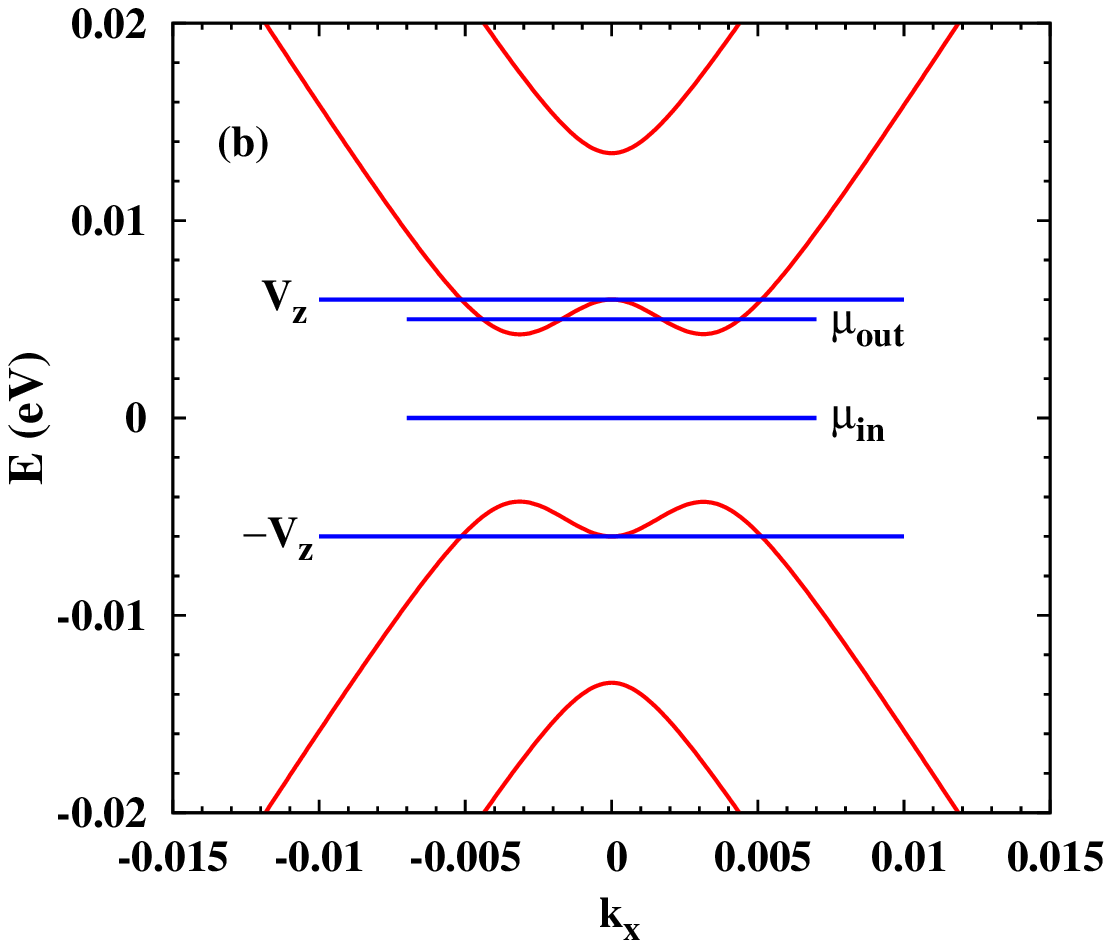}
\caption{(Color online) (a) Topological phase diagram in the ($\mu,V_z$) space
  with $\Delta\ne 0$ or $\Delta=0$. The dashed curves, i.e., $V^2_z=\Delta^2+\mu^2$ are the phase boundaries between
  the TSC and NTSC whereas the dotted ones, i.e, $\mu^2=E_0^2$ stand for the
  phase boundaries between the QAH state and metal. (b) Bulk energy spectrum of
  $H_{e}^{\rm eff}$ near the $K$ point with $k_y=0$ and $\Delta=0$. $V_z$ ($-V_z$) is the upper (lower) limit of the chemical
  potential in the topological nontrivial regime ($V_z^2>\mu^2+\Delta^2$). $\mu_{\rm in}$ and $\mu_{\rm out}$ stand for the chemical 
  potential in and out of the gap, respectively. $V_z=6\ $meV and $\lambda=4\ $meV.}
\label{fig1}
\end{figure}

In the following, we focus on the investigation near the
Dirac points. We first study the topological phase diagram as shown in
Fig.~\ref{fig1}(a). The phase boundaries between the topological and 
nontopological superconductors (NTSCs) are determined by the dashed curves,
i.e., $V_z^2=\mu^2+\Delta^2$. To
further distinguish the TSCs (i.e., $V_z^2>\mu^2+\Delta^2$), we suppress
the $s$-wave superconductivity. Without the $s$-wave superconductivity, we
show the bulk energy spectrum of the low energy effective Hamiltonian near the
Dirac points $H_e^{\rm eff}$ (see Appendix~\ref{appA}) in Fig.~\ref{fig1}(b). When the chemical potential
lies in the gap ($|\mu|<E_0$), eg., $\mu_{\rm in}$, the
system behaves as a QAH state with the Chern number $|N|=2$.\cite{qiao} 
Note that $E_0$ is the absolute value of the minimum (maximum)
energy of the conduction (valence) band with the formula given in Appendix~\ref{appA}. This QAH state in proximity to an $s$-wave
superconductor becomes a TSC with the Chern number $2|N|=4$ \cite{qi2}
(see regime I). When the chemical potential is tuned out of the
gap below the upper limit $|V_z|$, eg., ${\mu_{\rm out}}$, the system is in a metallic phase with two Fermi surfaces in each
valley as shown in Fig.~\ref{fig1}(b) ($K^{\prime}$ valley is not shown
here). With the $s$-wave superconductivity included, the
effective paring near each of these four Fermi surfaces is equivalent
to that of a $p$-wave superconductor.\cite{sau,alicea2,jianli} Each of
these effective $p$-wave superconductors hosts a Majorana edge mode,
which is in agreement with the Chern number near the Dirac
points, i.e., $|C_1|=4$. This effective $p$-wave superconductor from metal is
labeled as regime II. Similarly, the NTSCs (i.e., $V_z^2<\mu^2+\Delta^2$) can also be divided into two
regimes, i.e., regime III (from the QAH state) and regime IV (from metal).

\begin{figure}[bth]
\centering
\includegraphics[width=4.2cm]{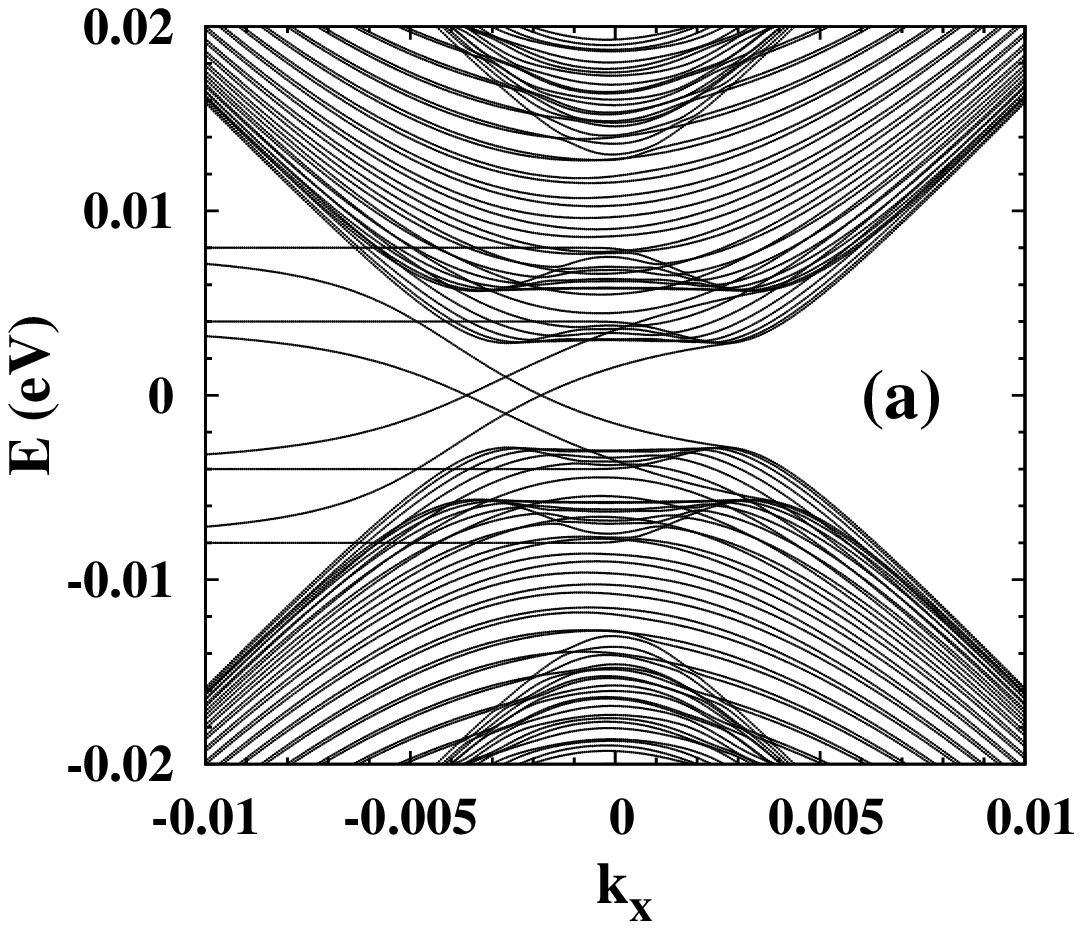}
\includegraphics[width=4.2cm]{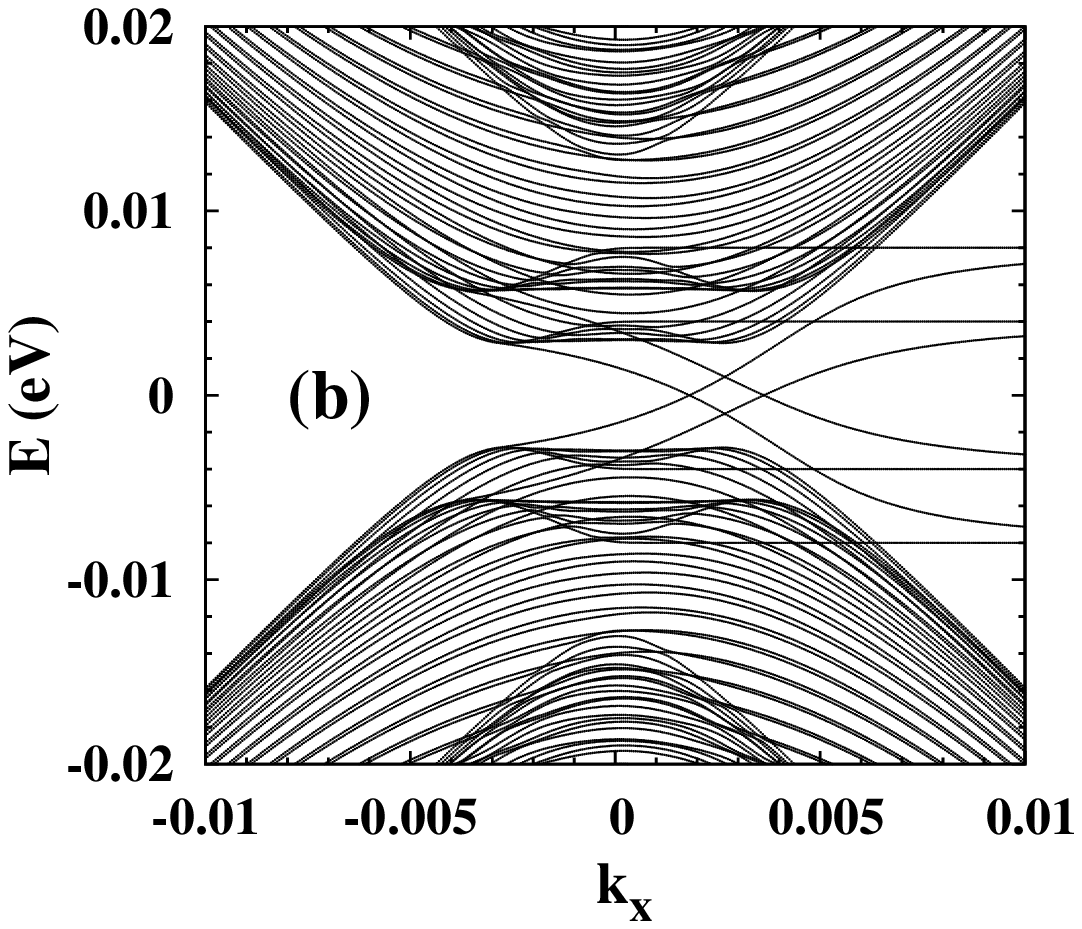}
\includegraphics[width=4.2cm]{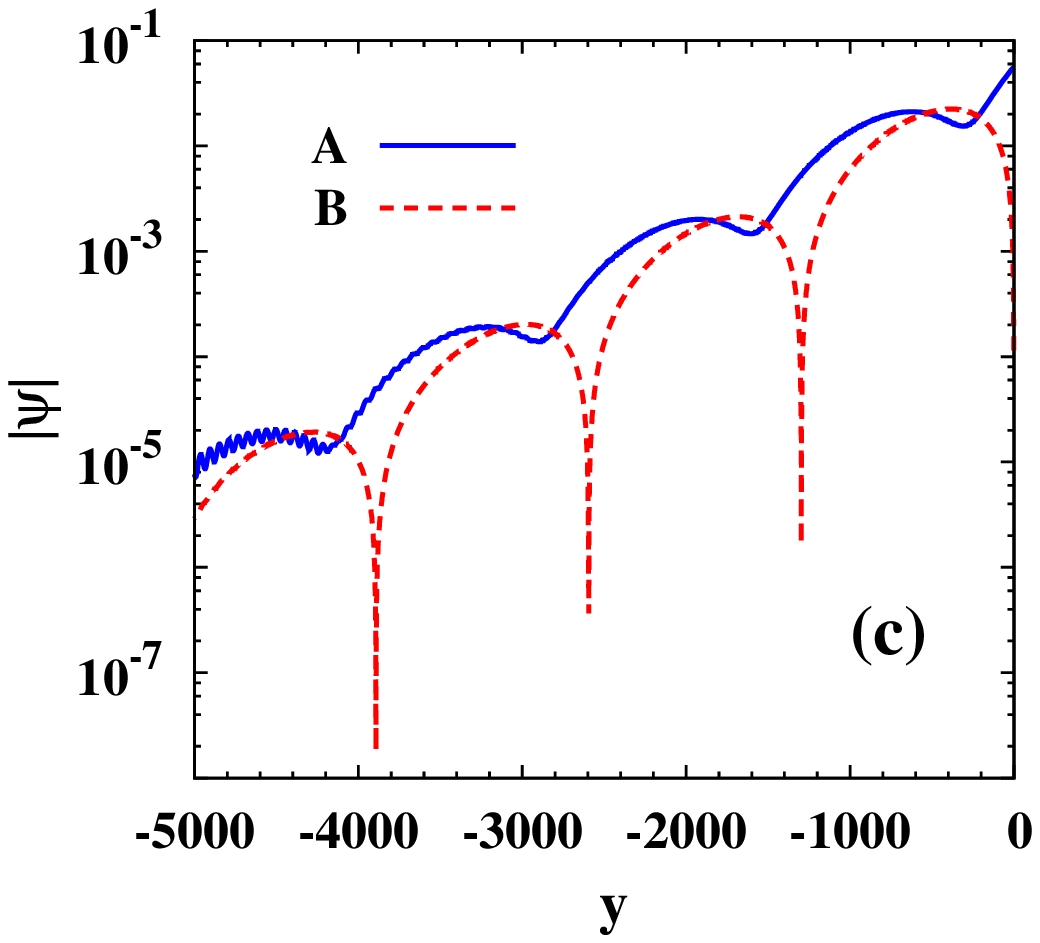}
\includegraphics[width=4.2cm]{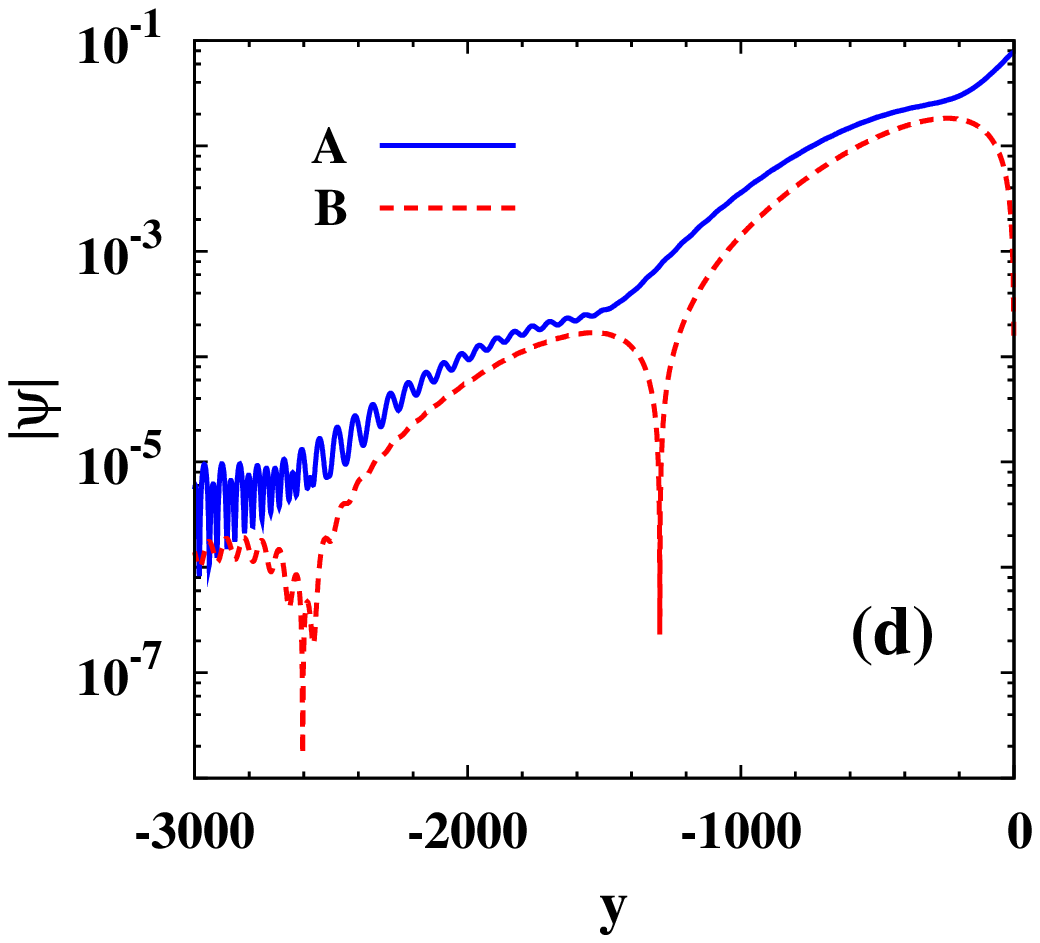}
\caption{(Color online) (a) and (b) represent the energy spectrum of zigzag
  graphene ribbon with the Rashba SOC, exchange field and proximity-induced $s$-wave
  superconductivity near the $K$ and $K^{\prime}$ points, respectively. (c) ((d))
  Real space probability amplitude $|\psi|$ across the width for the Majorana edge state 
with a smaller (larger) $|k_x|$ near the $K$ point at one edge (i.e., $y=0$) with
$v_x>0$ (only part of the ribbon is shown). A (B) refers to A (B) sublattice. The fluctuations of $|\psi|$ at the
positions far away from the edge are due to numerical error. Here, 
$V_z=6\ $meV, $\lambda=4\ $meV, $\mu=0$ and $\Delta=2\ $meV. 
}
\label{fig2}
\end{figure}

\subsection{Majorana edge states}
As the effective $p$-wave superconductors (regime II) have been
widely investigated in the
literature,\cite{sau,sau2,alicea2,lutchyn,alicea3,halperin,stanescu,mourik,yzhou,bysun}
we concentrate on the TSC from the QAH
state (regime I). The Majorana edge states are studied in thick
graphene ribbons. The numerical
method is detailed in Appendix~\ref{appC}. We plot the energy spectrum of zigzag
graphene ribbon near the $K$ and $K^{\prime}$ points in Figs.~\ref{fig2}(a) and (b), respectively. We find that 
there exist four zero energy states in each valley. These eight states can be
divided into two categories, i.e., four propagate along the same
direction $+x$ ($-x$) determined by the group 
velocity $v_x=\frac{1}{\hbar}\frac{\partial E(k_x)}{\partial k_x}$ $>0$
($<0$). Moreover, the four states in the same category are at the same edge, which is
in agreement with the magnitude of the Chern number, i.e., $|C_1|=4$. This
indicates that these eight zero energy states are topologically protected
Majorana edge states. Specifically, we choose two of them at the same edge with $v_x>0$ near the
$K$ point and show the real space probability amplitude of the one with smaller
and larger $|k_x|$ in Figs.~\ref{fig2}(c) and (d), respectively. Note that we separate the A and B sublattices by
the blue solid and red dashed curves. It is seen that the amplitudes
of both A and B sublattices in two Majorana edge states show obvious decay and
oscillation. However, the penetration lengths are different between
these two Majorana edge states.

\begin{figure}[bth]
\centering
\includegraphics[width=8.5cm]{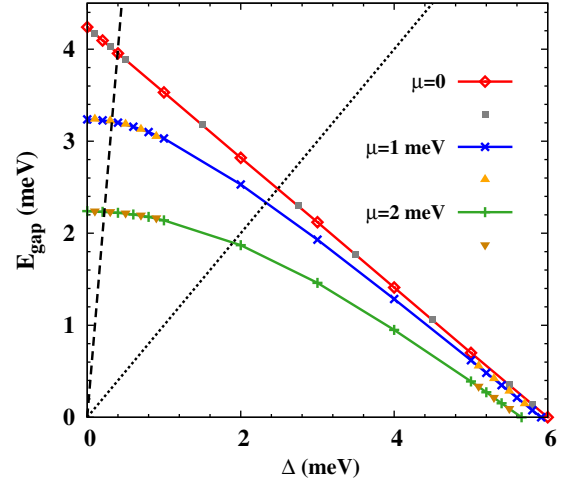}
\caption{(Color online) Bulk excitation gap $E_{\rm gap}$ of the TSC from the QAH state as a function of $\Delta$. The solid curves with
  diamonds, crosses and plus signs correspond to the numerical results at $\mu=0$,
  $1\ $meV and $2\ $meV, respectively. The analytical results at $\mu=0$, $1\
  $meV and $2\ $meV are separately represented by the symbols of squares, upward triangles and
  downward triangles. Note that for the analytical results at $\mu\ne 0$, only
  two limits, i.e., $\Delta\sim 0$ and $\Delta\sim \Delta_c$ are calculated. In addition, the dotted (dashed) curve
  corresponds to $E_{\rm gap}=\Delta$ ($E_{\rm gap}=10\Delta$). $V_z=6\ $meV and
  $\lambda=4\ $meV.}
\label{fig3}
\end{figure}

\subsection{Bulk excitation gap}
\subsubsection{Chemical potential dependence}
The above Majorana edge states are protected by a bulk
excitation gap of the TSC from the QAH state. With different chemical potentials
chosen in the gap of a QAH system, the bulk excitation gap 
as a function of the proximity-induced superconducting gap is plotted in
Fig.~\ref{fig3}. In the $\Delta=0$ limit, the system can be considered as
  two copies of QAH insulators as shown in Fig.~\ref{fig1}(b) but with an energy
  shift of $-\mu$ ($\mu$) for the particle (hole) one. Then, the bulk excitation gap
  of our system is determined by these two QAH insulators, i.e., $E_{\rm
    gap}=E_{0}-|\mu|$. This nonzero bulk excitation gap in the limit $\Delta=0$
  strongly indicates that the bulk excitation gap can be
much larger than $\Delta$ especially for small $\Delta$. It is emphasized that the nonzero excitation gap in the $\Delta=0$
limit is totally different from
the case of the effective $p$-wave superconductors where the excitation gap is
exactly zero in the limit $\Delta=0$.\cite{alicea2} At the critical point
$\Delta_c=\sqrt{V_z^2-\mu^2}$, the bulk excitation gap of our system becomes zero. 
In between, the bulk excitation gap shows a monotonic decrease with increasing $\Delta$. We emphasize that during this process, $E_{\rm
  gap}$ can be larger or even one order of magnitude larger than $\Delta$ by
referring to $E_{\rm gap}=\Delta$ (dotted curve) and $E_{\rm
  gap}=10\Delta$ (dashed curve). For example, $E_{\rm gap}=4.02\ $meV
($\Delta=0.3\ $meV) at $\mu=0$; $E_{\rm gap}=3.22\ $meV ($\Delta=0.3\ $meV) at
$\mu=1\ $meV; $E_{\rm gap}=2.23\ $meV ($\Delta=0.2\ $meV) at $\mu=2\ $meV. This marked enlargement of
the gap is in strong
contrast to the effective $p$-wave superconductors where the bulk excitation gap is always smaller than
the induced superconducting gap.\cite{sau,sau2,alicea2,lutchyn,alicea3,halperin,stanescu,mourik,yzhou,bysun} This makes our proposal, 
i.e., the TSC from the QAH state, very promising for the realization of robust
Majorana modes in experiments.    

To have a better understanding of the behavior of the bulk excitation
  gap of the TSC from the QAH state, we also perform an analytic derivation. Near the Dirac points,
  the BdG Hamiltonian $H_{\rm BdG}({\bf k})$ in Eq.~(3) can be expanded as a low energy effective one
  with $H_e({\bf k})$ [see Eq.~(\ref{hamilnosc})] being replaced by $H^{\rm
    eff}_{e}({\bf k})$ [see Eq.~(\ref{effhamil})]. The secular equation of the
  eigenvalue $E$ is ${\rm det}[H_{\rm BdG}({\bf k})-EI_{8\times 8}]=0$ where
  $I_{8\times 8}$ is a unit matrix. After a careful calculation, we have 
\begin{eqnarray}
&&[\alpha_1^2-4V_z^2\alpha_3+4\alpha_1 (\lambda_R^2-\mu^2-\mu
V_z)+4\alpha_2(\mu^2-\lambda_R^2)]^2\nonumber\\
&&\mbox{}-64V_z^2\alpha_3(\lambda_R^2-\mu^2-\mu
V_z)^2+8[\alpha_1\mu-2(\mu+V_z)\nonumber\\
&&\mbox{}\times(\mu^2-\lambda_R^2)][(\mu+V_z)(\alpha_1^2-4V_z^2\alpha_3)-2\mu
\alpha_1\alpha_2]=0\nonumber\\\label{secular}
\end{eqnarray}
with $\alpha_1=\alpha_2-\alpha_3+\alpha_4$, $\alpha_2=v_f^2k_x^2$,
$\alpha_3=E^2$, $\alpha_4=\Delta^2-V_z^2+\mu^2$, $v_f=3t/2$ and
$\lambda_R=3\lambda/2$. Note that we focus on the calculation near the $K=(4\pi/3,0)$ ($\tau=1$)
and set $k_y=0$ by considering the isotropy of the low energy effective
Hamiltonian. It is very difficult to obtain the eigenvalues by solving
Eq.~(\ref{secular}) directly. Instead of the eigenvalues, we are
interested in the bulk excitation gap here. Differentiating
Eq.~(\ref{secular}) with respect to $\alpha_2$ and then employing the extreme
value condition of the excitation gap (i.e., $\frac{\partial
  \alpha_3}{\partial \alpha_2}=0$), we have
\begin{eqnarray}
\alpha_3^3-g_2\alpha_3^2-g_1\alpha_3-g_0=0\label{diff}
\end{eqnarray}
where $g_2=3(\alpha_2+\alpha_4)+2(2\lambda_R^2-\mu^2+2V_z^2)$,
$g_1=-3(\alpha_2+\alpha_4)^2+4(-2\lambda_R^2-V_z^2+\mu^2)(\alpha_2+\alpha_4)+4\alpha_2(\lambda_R^2+\mu^2)-8V_z^2(\lambda_R^2+\mu^2)$
and $g_0=(\alpha_2+\alpha_4)^3-2(\mu^2-2\lambda_R^2)(\alpha_2+\alpha_4)^2+4(\alpha_2+\alpha_4)
[\alpha_2(-\lambda_R^2-\mu^2)-2\lambda_R^2(\mu^2-V_z^2-\lambda_R^2)]+8\alpha_2(\mu^4-\lambda_R^4)
+8\alpha_4\lambda_R^2(\mu^2-\lambda_R^2)$. 

At $\mu=0$, Eq.~(\ref{diff}) can be simplified to $(\alpha_2+\alpha_4-\alpha_3)(4\alpha_3^2+q_1\alpha_3+q_2)=0$
with $q_1=-8(\alpha_2+\alpha_4)-16(\lambda_R^2+V_z^2)$ and
$q_2=4(\alpha_2+\alpha_4)^2+16\lambda_R^2(\alpha_2+\alpha_4)-16\lambda_R^2\alpha_2+32\lambda_R^2V_z^2$. Since
the equation $4\alpha_3^2+q_1\alpha_3+q_2=0$ is inconsistent with the gap
closing condition, we only have $\alpha_2+\alpha_4-\alpha_3=\alpha_1=0$. With this condition together with Eq.~(\ref{secular}), one
obtains the bulk excitation gap $E_{\rm gap}=E_0(1-|\Delta|/|V_z|)$, which is linearly dependent on $\Delta$ and agrees very well
with the numerical results as shown in Fig.~\ref{fig3}. Specially, for $E_{\rm
  gap}>|\Delta|$ ($E_{\rm gap}>10|\Delta|$), we have $|\Delta|<E_0|V_z|/(E_0+|V_z|)\equiv \Delta_1$
($|\Delta|<E_0|V_z|/(E_0+10|V_z|)\equiv \Delta_2\approx 0.1E_0$). These conditions will guide the
experiments to obtain robust TSCs and Majorana modes. As for the case of $\mu\ne 0$, it is very difficult for us
to obtain an exact analytic solution. Only the analytical results in two limits,
i.e., $|\Delta|\sim \Delta_c$ and $\Delta\sim 0$, are given. In the $|\Delta|\sim
\Delta_c$ limit, we have $E_{\rm
  gap}=(\Delta_c-|\Delta|)\Delta^2_c|\mu^2-\lambda_R^2|/\sqrt{{V_z^2(\Delta_c^2+\lambda_R^2)(\mu^4+\lambda_R^2\Delta_c^2)}}$. In
the limit $\Delta\sim 0$, $E_{\rm
gap}=\sqrt{(E_0-\mu)^2-\Delta^2w_2/w_1}$ with
$w_1=-\lambda_R^2\mu^2V_z^2+\mu (\lambda_R^2+V_z^2)(-\lambda_R^2+\mu^2-V_z^2)E_0$ and
$w_2=\mu (\lambda_R^2-\mu^2)(\lambda_R^2+V_z^2)E_0-16\lambda_R^2V_z^2(\lambda_R^4-\mu^2V_z^2-2\lambda_R^2\mu^2
+\lambda_R^2V_z^2)/(\lambda_R^2+V_z^2)$
by assuming $0<\mu<E_0$. The analytical results at $\mu\ne 0$ in both limits agree fairly well with the
numerical ones as shown in Fig.~3.  

\begin{figure}[bth]
\centering
\includegraphics[width=8.5cm]{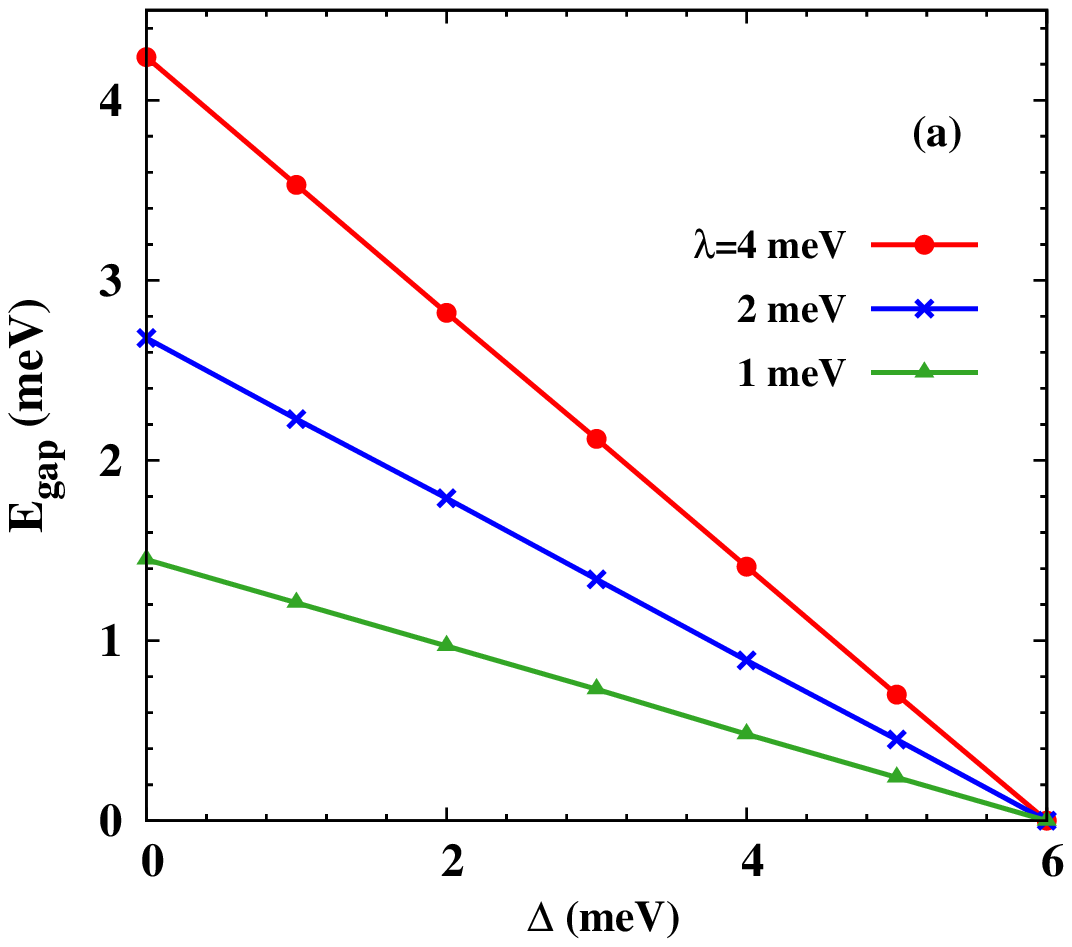}
\includegraphics[width=8.5cm]{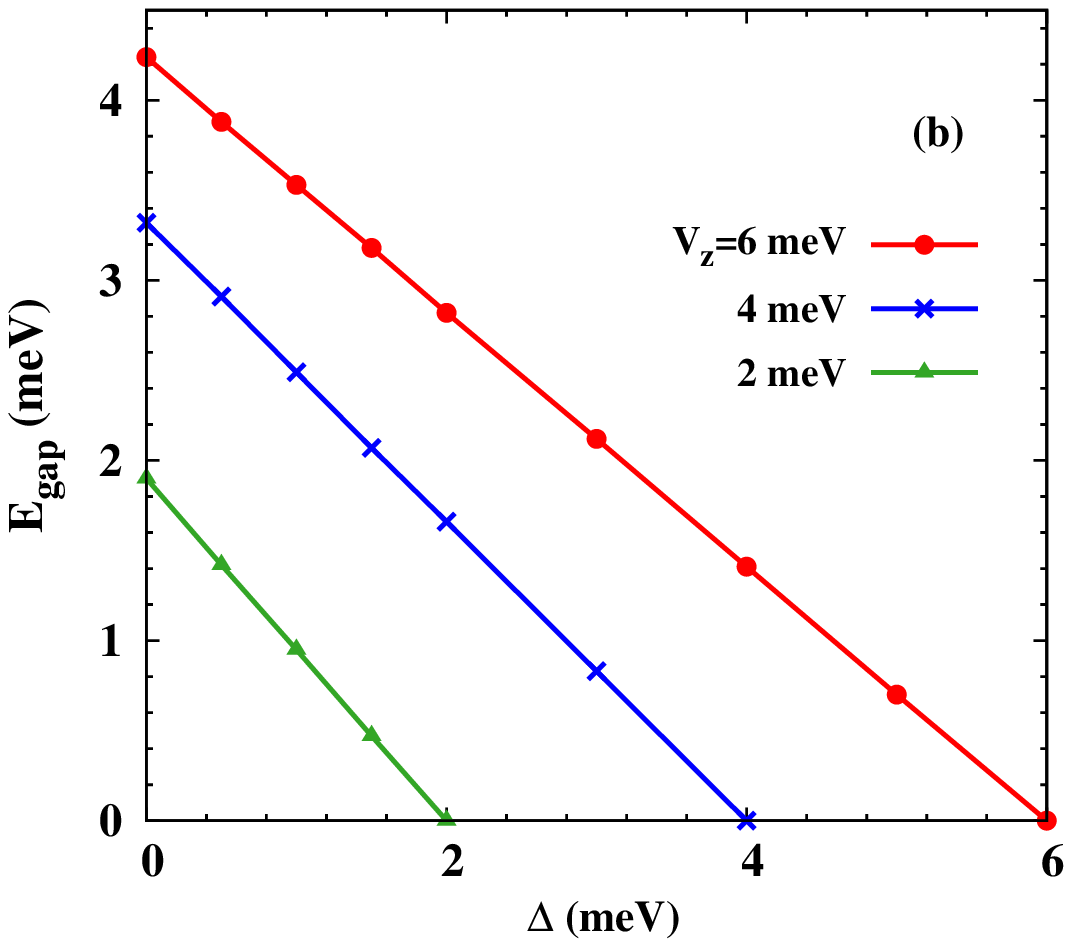}
\caption{(Color online) Numerical results of bulk excitation gap $E_{\rm gap}$ of the TSC from the QAH
  state as a function of the proximity-induced superconducting gap $\Delta$ at
  $\mu=0$ (a) under different $\lambda$ with $V_z=6\ $meV and (b) under
  different $V_z$ with $\lambda=4\ $meV.}
\label{S2}
\end{figure}

\subsubsection{Rashba SOC strength and exchange field dependences}
We then turn to investigate the effects of the Rashba SOC and exchange
field on the bulk excitation gap of the TSC
from the QAH state. In Figs.~\ref{S2}(a) and (b), we plot the
dependence of the bulk excitation gap on the proximity-induced superconducting
gap at $\mu=0$ under different Rashba SOC strengths and exchange fields, respectively. 
It is seen that the bulk excitation gap increases with the increase of either the
Rashba SOC strength or exchange field. This can be easily understood from $E_{\rm gap}=E_0(1-|\Delta|/|V_z|)$ mentioned above where $E_0$ (see Appendix~\ref{appA}) 
increases with increasing Rashba SOC strength and exchange field.

\section{SUMMARY AND DISCUSSION}
In summary, we have proposed that in the presence of proximity-induced
$s$-wave superconductivity, the QAH state due to the Rashba SOC and exchange field
in single layer graphene can become a two-dimensional TSC. With very low or
even zero doping near the Dirac points, i.e., two inequivalent valleys, we show 
that this TSC, which exhibits a Chern number as large as
four and hosts four Majorana edge modes, has a bulk excitation gap being
larger or even one order of magnitude lager than the proximity-induced
superconducting gap. The unique feature is in strong contrast to the case of
the effective $p$-wave superconductors where the bulk excitation gap is always
smaller than the proximity-induced superconducting gap. This also applies
  to other QAH systems as seed materials to obtain robust TSCs and Majorana modes.

Finally, we address the experimental feasibility of the TSC from the QAH
state. Single layer
graphene on the (111) surface of an antiferromagnetic insulator BiFeO$_3$ can
have an exchange field ($V_z=142\ $meV) and Rashba SOC ($\lambda=1.4\ $meV),
realizing a QAH insulator with a gap being $2E_0=4.2\ $meV.\cite{qiao2} 
This QAH state ($|\mu|<E_0$) in proximity to a conventional $s$-wave superconductor (eg., Nb with a large
superconducting gap $\Delta_{\rm Nb}=0.83\ $meV \cite{jwang}) becomes a TSC since
the topologically nontrivial condition $\Delta^2+\mu^2<V_z^2$ is easily satisfied due
to $|\mu|<E_0\ll |V_z|$ and $|\Delta|<|\Delta_{\rm Nb}|\ll |V_z|$. With $\Delta=0.5\ $meV ($\Delta_2=0.21\ $meV
  $<\Delta<\Delta_1=2.1\ $meV) for estimation, we have the bulk excitation gap
$E_{\rm gap}=2.05\ $meV, $1.56\ $meV and $1.06\ $meV, corresponding to a
temperature of $23.8\ $K, $18.1\ $K and $12.3\ $K, at $\mu=0$, $0.5\ $meV and
$1\ $meV, respectively. The large excitation gap (of the order of $10\ $K)
ensures that robust Majorana modes can be achieved.

\begin{acknowledgments}
This work was supported by the National Natural Science Foundation of
China under Grant No.\ 11334014 and 61411136001, the National Basic 
Research Program of China under Grant No.\ 2012CB922002 and the Strategic 
Priority Research Program of the
Chinese Academy of Sciences under Grant No.\ XDB01000000. 
\end{acknowledgments}

\begin{appendix}
\section{$H_e({\bf k})$ in Eq.~(3) near the Dirac points}\label{appA}
Near the Dirac points, i.e., $K=(4\pi/3,0)\ (\tau=1)$ and $K^{\prime}=(-4\pi/3,0)\ (\tau=-1)$,
$H_e({\bf k})$ in Eq.~(3) can be expanded as a low energy effective Hamiltonian 
\begin{widetext}
\begin{eqnarray}
H^{\rm eff}_{e}({\bf k})=\left(\begin{array}{cccc}
V_z & v_f(\tau k_x-ik_y) & 0 & i\lambda_R(1-\tau)\\
v_f(\tau k_x+ik_y) & V_z & -i\lambda_R(1+\tau) & 0\\
0 & i\lambda_R(1+\tau) & -V_z & v_f(\tau k_x-ik_y)\\
i\lambda_R(\tau-1) & 0 & v_f(\tau k_x+ik_y) & -V_z\label{effhamil}
\end{array}\right).
\end{eqnarray}
\end{widetext}
The energy spectrum of this
effective Hamiltonian is shown in Fig.~1(b). The minimum (maximum) energy of the
conduction (valence) band is $E_0=|V_z\lambda_R|/\sqrt{V_z^2+\lambda_R^2}$
($-E_0$) after a simple calculation and then the band gap is given by $2E_0$.  

\section{Gap closing condition of the BdG Hamiltonian $H_{\rm BdG}({\bf k})$}\label{appB}
The gap of $H_{\rm BdG}({\bf k})$ closes at the momenta $\Gamma$ (single one), $M$ (three
inequivalent ones) and $K$ (two inequivalent ones) points. Specifically, at the
momentum $\Gamma$, the Rashba SOC vanishes [see Eq.~(\ref{hamilnosc})], which is similar to the previous
studies in semiconductors.\cite{sau,sau2} The gap closing condition is given by $(\mu\pm
3t)^2=V^2_z-\Delta^2$ with $+$ ($-$) representing lower (higher) energy band at
$\Gamma$ after a simple calculation. As for the momentum $M$, the
Rashba SOC does not cause spin splitting but lead to an energy shift for the
spin degenerate bands. We take $M=(0,\frac{2\sqrt{3}\pi}{3})$ for example and 
$H_{\rm BdG}(M)$ [see Eq.~(3)] reads
\begin{widetext}
\begin{eqnarray}
H_{\rm BdG}(M)=\left(\begin{array}{cccccccc}
-\mu+V_z & \frac{\sqrt{3}i-1}{2}t & 0 & -\lambda(i+\sqrt{3}) & \Delta & 0 & 0 & 0\\
\frac{-\sqrt{3}i-1}{2}t & -\mu+V_z & -\lambda(\sqrt{3}-i) & 0 & 0 & \Delta & 0 & 0\\
0 & -\lambda(i+\sqrt{3}) & -\mu-V_z & \frac{\sqrt{3}i-1}{2}t & 0 & 0 & \Delta & 0\\
-\lambda(\sqrt{3}-i) & 0 & \frac{-\sqrt{3}i-1}{2}t & -\mu-V_z & 0 & 0 & 0 & \Delta\\
\Delta & 0 & 0 & 0 & \mu+V_z & -\frac{\sqrt{3}i-1}{2}t & 0 & \lambda(i+\sqrt{3})\\
0 & \Delta & 0 & 0 & \frac{\sqrt{3}i+1}{2}t & \mu+V_z & \lambda(\sqrt{3}-i) & 0\\
0 & 0 & \Delta & 0 & 0 & \lambda(i+\sqrt{3}) & \mu-V_z & -\frac{\sqrt{3}i-1}{2}t\\
0 & 0 & 0 & \Delta & \lambda(\sqrt{3}-i) & 0 & \frac{\sqrt{3}i+1}{2}t & \mu-V_z 
\end{array}\right).
\end{eqnarray} 
Performing a unitary transformation as $\tilde{H}_{\rm
  BdG}(M)=U_M^{\dagger}H_{\rm BdG}(M)U_M$ with
\begin{eqnarray}
U_M=\frac{\sqrt{2}}{2}\left(\begin{array}{cccccccc}
0 & \frac{1-\sqrt{3}i}{2} & 0 & \frac{-1+\sqrt{3}i}{2} & 0 & 0 & 0 & 0\\
0 & 1 & 0 & 1 & 0 & 0 & 0 & 0\\
\frac{1-\sqrt{3}i}{2} & 0 & \frac{-1+\sqrt{3}i}{2} & 0 & 0 & 0 & 0 & 0\\
1 & 0 & 1 & 0 & 0 & 0 & 0 & 0\\
0 & 0 & 0 & 0 & 0 & \frac{1-\sqrt{3}i}{2} & 0 & \frac{-1+\sqrt{3}i}{2}\\
0 & 0 & 0 & 0 & 0 & 1 & 0 & 1\\
0 & 0 & 0 & 0 & \frac{1-\sqrt{3}i}{2} & 0 & \frac{-1+\sqrt{3}i}{2} & 0\\
0 & 0 & 0 & 0 & 1 & 0 & 1 & 0 
\end{array}\right),
\end{eqnarray}
one obtains
\begin{eqnarray}
\tilde{H}_{\rm BdG}(M)=\left(\begin{array}{cccccccc}
-t-\mu-V_z & 0 & 0 & -2i\lambda & \Delta & 0 & 0 & 0\\
0 & -t-\mu+V_z & -2i\lambda & 0 & 0 & \Delta & 0 & 0\\
0 & 2i\lambda & t-\mu-V_z & 0 & 0 & 0 & \Delta & 0\\
2i\lambda & 0 & 0 & t-\mu+V_z & 0 & 0 & 0 & \Delta\\
\Delta & 0 & 0 & 0 & t+\mu-V_z & 0 & 0 & 2i\lambda\\
0 & \Delta & 0 & 0 & 0 & t+\mu+V_z & 2i\lambda & 0\\
0 & 0 & \Delta & 0 & 0 & -2i\lambda & -t+\mu-V_z & 0\\
0 & 0 & 0 & \Delta & -2i\lambda & 0 & 0 & -t+\mu+V_z 
\end{array}\right).
\end{eqnarray} 
At $\mu\sim t$, the block with the diagonal terms
being $-t-\mu\mp V_z$ and $t+\mu\mp V_z$ in $\tilde{H}_{\rm BdG}(M)$ is far from gap closing whereas the
gap closing is determined by the remaining one. By considering that $|\lambda|\ll t$,
we use the L\"{o}wdin partition method \cite{lowdin,winkler} to obtain the effective 
Hamiltonian for the block determining the gap closing as 
\begin{eqnarray}
H_{\rm eff}(M)=\left(\begin{array}{cccc}
t-\mu-V_z+\frac{2\lambda^2}{t-V_z} & 0 & \Delta & 0\\
0 & t-\mu+V_z+\frac{2\lambda^2}{t+V_z} & 0 & \Delta\\
\Delta & 0 & -t+\mu-V_z-\frac{2\lambda^2}{t+V_z} & 0\\
0 & \Delta & 0 & -t+\mu+V_z+\frac{2\lambda^2}{-t+V_z}
\end{array}\right).
\end{eqnarray}
Then, the gap closing condition is
$(t-\mu-V_z+\frac{2\lambda^2}{t-V_z})(-t+\mu-V_z-\frac{2\lambda^2}{t+V_z})-\Delta^2=0$
or
$(t-\mu+V_z+\frac{2\lambda^2}{t+V_z})(-t+\mu+V_z-\frac{2\lambda^2}{-t+V_z})-\Delta^2=0$. 
As $|V_z|\ll t$, both conditions become
$(t-\mu+\frac{2\lambda^2}{t})^2=V_z^2-\Delta^2$ approximately. Furthermore, by
considering that $|\lambda|\ll t$, we neglect the energy shift of
$2\lambda^2/{t}$ and then the gap closing condition at the momentum $M$
with $\mu\sim t$ is given by $(t-\mu)^2=V_z^2-\Delta^2$. Similarly, the gap closing
condition at $M$ with $\mu\sim -t$ is $(t+\mu)^2=V_z^2-\Delta^2$ under the
approximation $|\lambda|,|V_z|\ll t$.

In contrast to the momenta $\Gamma$ and $M$,  the Rashba SOC at the Dirac points
contributes to a finite spin splitting. Specifically, with $K=(4\pi/3,0)$,
$H_{\rm BdG}(K)$ [see Eq.~(3)] can be
written as
\begin{eqnarray}
H_{\rm BdG}({K})=\left(\begin{array}{cccccccc}
-\mu+V_z & 0 & 0 & 0 & \Delta & 0 & 0 & 0\\
0 & -\mu+V_z & -3i\lambda & 0 & 0 & \Delta & 0 & 0\\
0 & 3i\lambda & -\mu-V_z & 0 & 0 & 0 & \Delta & 0\\
0 & 0 & 0 & -\mu-V_z & 0 & 0 & 0 & \Delta\\
\Delta & 0 & 0 & 0 & \mu+V_z & 0 & 0 & 0\\
0 & \Delta & 0 & 0 & 0 & \mu+V_z & 3i\lambda & 0\\
0 & 0 & \Delta & 0 & 0 & -3i\lambda & \mu-V_z & 0\\
0 & 0 & 0 & \Delta & 0 & 0 & 0 & \mu-V_z\\ 
\end{array}\right),
\end{eqnarray}  
which can be divided into two independent $4\times 4$ parts, i.e., 
$H_1$ ($H_2$) without (with) the Rashba SOC terms. Then, we have 
\begin{eqnarray}
H_1=\left(\begin{array}{cccc}
-\mu+V_z & 0 & \Delta & 0\\
0 & -\mu-V_z & 0 & \Delta\\
\Delta & 0 & \mu+V_z & 0\\
0 & \Delta & 0 & \mu-V_z\\
\end{array}\right),
\end{eqnarray}
which is exactly the same as the Hamiltonian of semiconductors with the Rashba SOC,
magnetic field and proximity-induced $s$-wave superconductivity at the momentum
$\Gamma$.\cite{sau,sau2} This indicates that both have the same gap closing condition, i.e.,
$V^2_z=\mu^2+\Delta^2$.\cite{sau,sau2} As for $H_2$ (not shown), due to the existence of the nonzero Rashba
SOC terms, the gap is always opened. Therefore, the 
gap closing condition at $K$ is just the one in $H_1$ part. Similar
analysis can be applied to $K^{\prime}$ and we obtain the same gap closing
condition as $K$. 
\end{widetext}

\section{Numerical method for calculating Majorana edge states in
  zigzag and armchair graphene ribbons}\label{appC}
We investigate the Majorana edge states near the Dirac
points in both zigzag and armchair graphene ribbons. We first study the case of
zigzag configuration. The Hamiltonian of zigzag ribbon can be obtained from
Eq.~(1) by choosing a unit cell and performing a Fourier transformation along
the direction parallel to the edge (assuming $x$-direction). Note that the unit
cell of the zigzag ribbon is the same as the one in
Ref.~\onlinecite{brey}. Specifically, 
\begin{widetext}
\begin{eqnarray}
H_{\rm zigzag}&=&-t\sum_{k_x}\sum_{\langle j_1,j_2
  \rangle\sigma}[1+|{\rm sgn}(x_{j_2}-x_{j_1})|e^{ik_x {\rm
    sgn}(x_{j_2}-x_{j_1})}]c^{\dagger}_{k_xj_1\sigma}c_{k_xj_2\sigma}+\sum_{k_x}\sum_{j\sigma}(\sigma
V_z-\mu)c^{\dagger}_{k_xj\sigma}c_{k_xj\sigma}\nonumber\\
&&\mbox{}+\Delta\sum_{k_x}\sum_{j}(c^{\dagger}_{k_xj\uparrow}c^{\dagger}_{-k_xj\downarrow}+{\rm
  H.c.})+i\lambda\sum_{k_x}\sum_{\langle j_1,j_2
  \rangle\sigma\sigma^{\prime}}[(\sigma_x^{\sigma\sigma^{\prime}}d^y_{j_1j_2}-\sigma_y^{\sigma\sigma^{\prime}}d^x_{j_1j_2})
+|{\rm sgn}(x_{j_2}-x_{j_1})|e^{ik_x{\rm sgn}(x_{j_2}-x_{j_1})}\nonumber\\
&&\mbox{}\times(\sigma_x^{\sigma\sigma^{\prime}}d^y_{j_1j_2}
+\sigma_y^{\sigma\sigma^{\prime}}d^x_{j_1j_2})]c^{\dagger}_{k_xj_1\sigma}c_{k_xj_2\sigma^{\prime}}
\end{eqnarray}
\end{widetext}
where $x_{j_2}-x_{j_1}$ is the relative position between $j_2$-th and $j_1$-th
atoms in the unit cell along the $x$-direction and sgn stands for the sign
function. By exactly diagonalizing $H_{\rm
  zigzag}$, one obtains the eigenvalues and eigenstates. However, this
method fails due to the computational limitations when the width of the ribbon becomes very large
(eg., of the order of $10^4$ atoms in the unit cell in our calculation). Alternatively, the zigzag
ribbon with the leading term, i.e., the hopping term, can be solved analytically near the
Dirac points.\cite{neto} Near $K$ ($\tau=1$) and $K^{\prime}$ ($\tau=-1$), the eigenstates are given by
\begin{eqnarray}
\Psi_{\tau k_x}^{z,\varepsilon}({\bf r})&=&Ae^{i(\tau|K|+k_x)x}\nonumber\\
&&\hspace{-1.6cm}\mbox{}\times\left(\begin{array}{c}
-v_f[(z-\tau k_x)e^{zy}+(z+\tau k_x)e^{-zy}]/\varepsilon \\
e^{zy}-e^{-zy} \\
\end{array}\right),\label{zig}
\end{eqnarray} 
with the eigenvalues being $\varepsilon^2=v_f^2(k_x^2-z^2)$ and
$A=\sqrt{\frac{\sqrt{3}}{|2(e^{2zL}-e^{-2zL})/z-8L|}}$. $L$ is the width of the
ribbon and $z$ is determined
by the equation $e^{-2zL}=(k_x+\tau z)/(k_x-\tau z)$. Note that if $z_0$ is a solution of
this equation, so does $-z_0$. As $\Psi_{\tau k_x}^{z_0,\varepsilon}=-\Psi_{\tau k_x}^{-z_0,\varepsilon}$, only one of
these two equivalent eigenstates needs to be taken. Then, one can use these eigenstates in Eq.~(\ref{zig}) with
additional spin and particle-hole degrees of freedom included to construct complete basis
functions for $H_{\rm zigzag}$. We diagonalize the Hamiltonian matrix of
$H_{\rm zigzag}$ and obtain the energy spectrum and
wavefunctions as shown in Fig.~2.

\begin{figure}[bth]
\centering
\includegraphics[width=6cm]{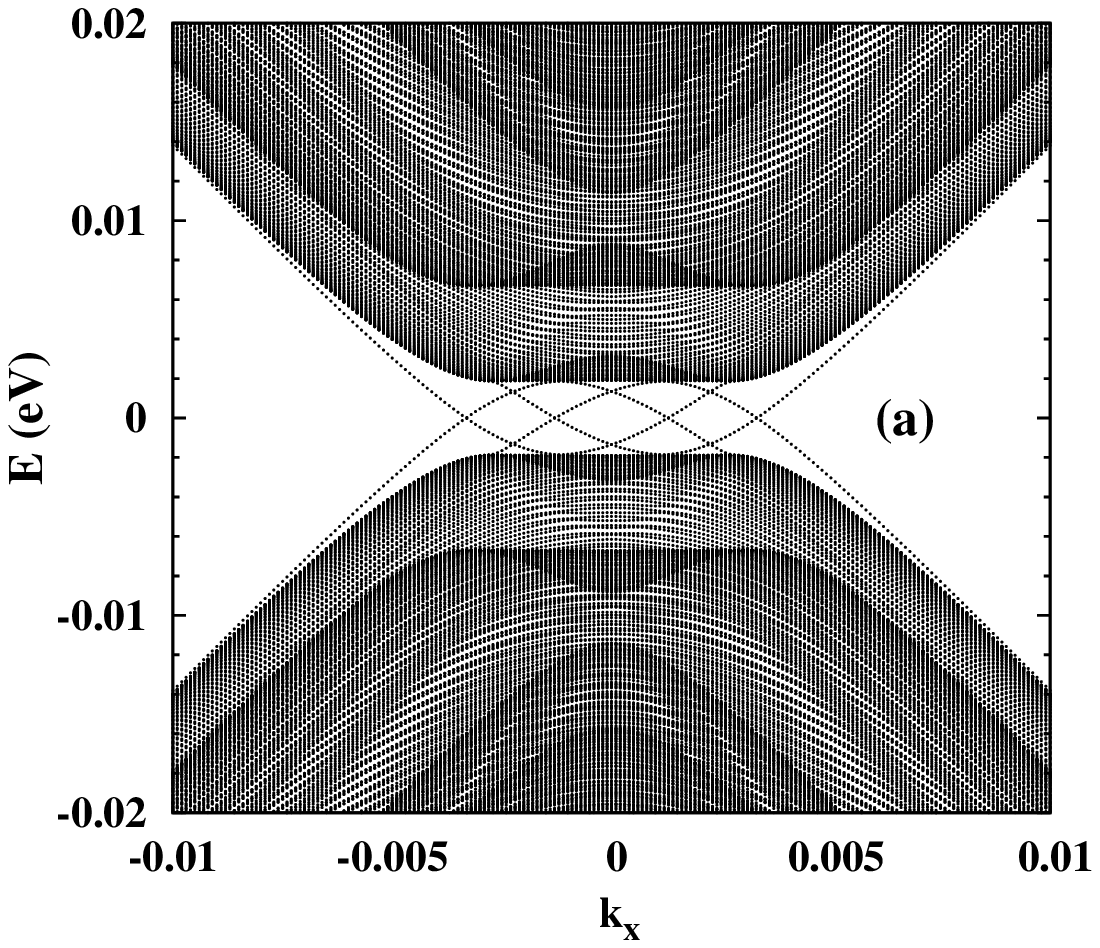}\\
\includegraphics[width=6cm]{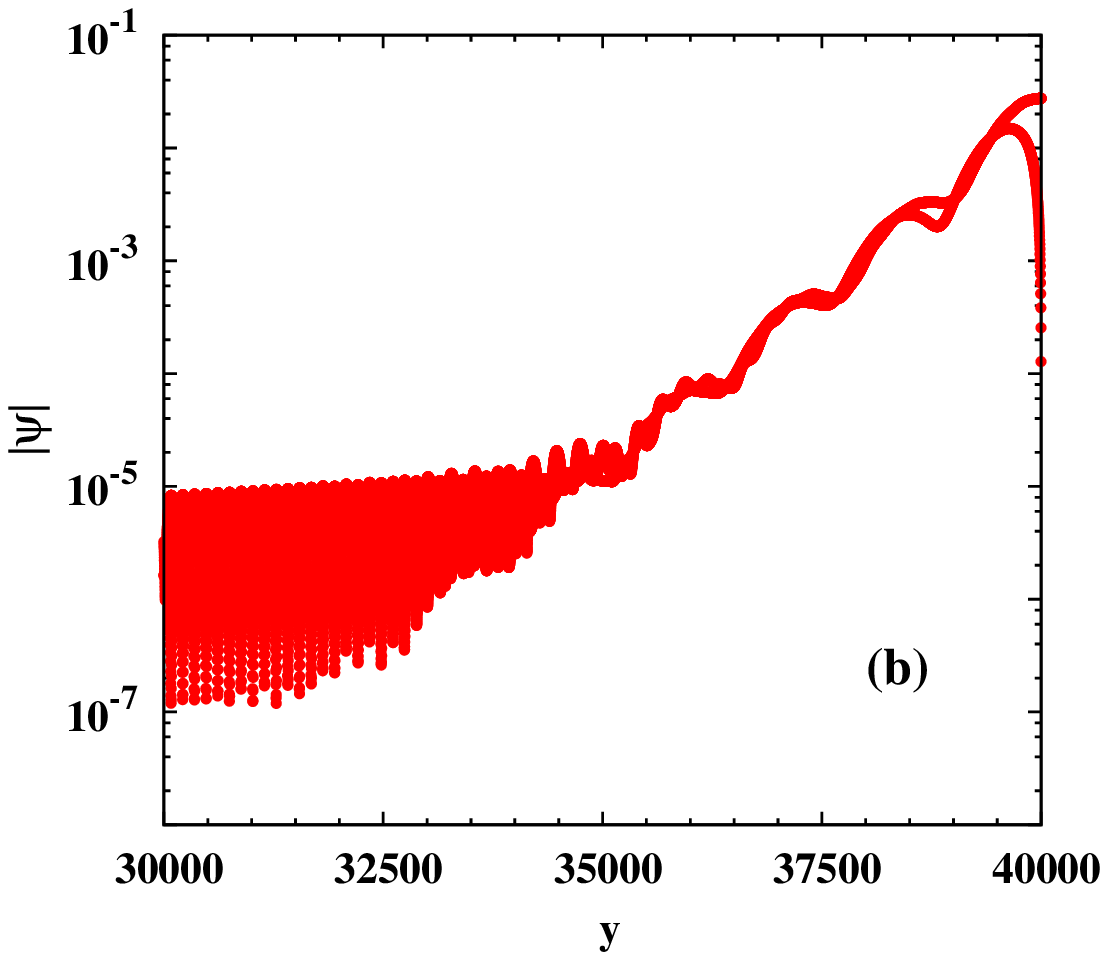}
\includegraphics[width=6cm]{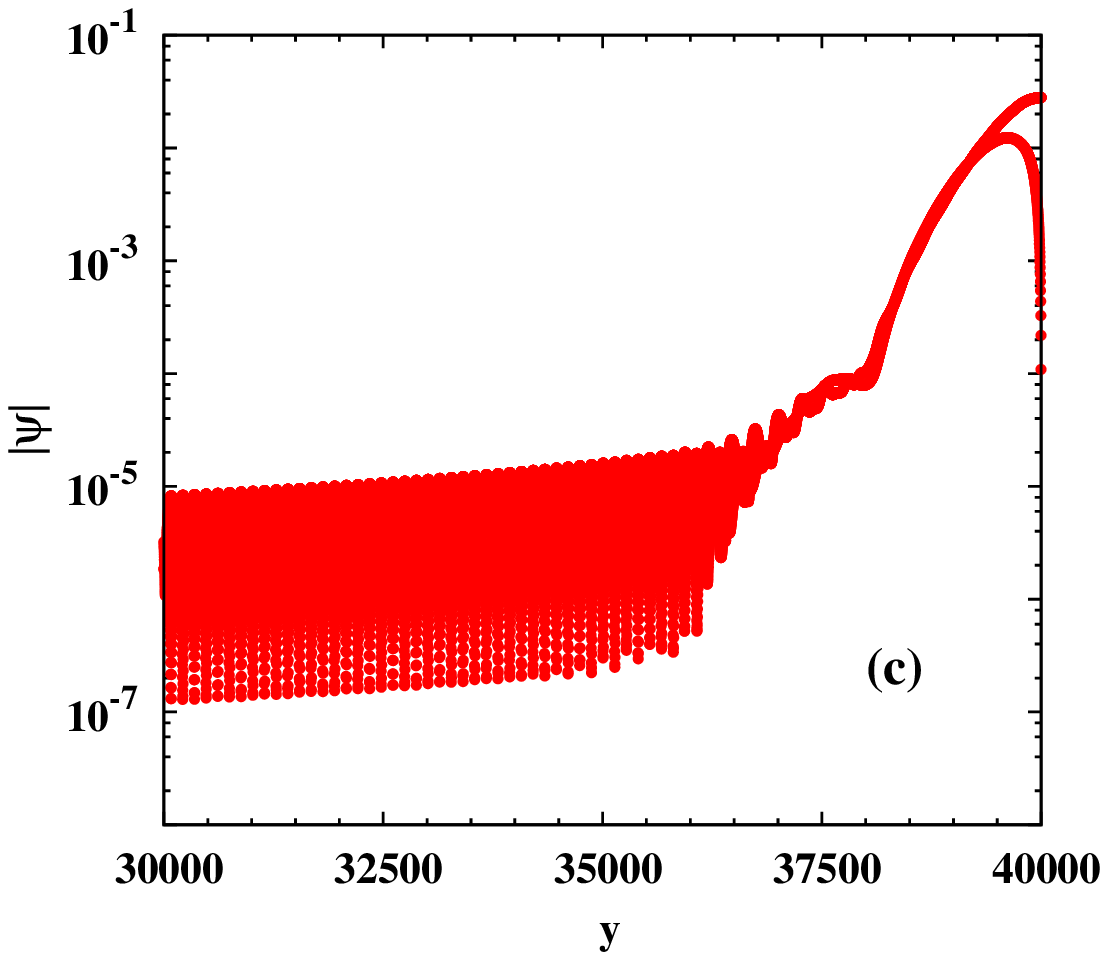}
\caption{(Color online) (a) Energy spectrum of armchair graphene ribbon in
  the presence of the Rashba SOC, exchange field and $s$-wave superconductivity from
  the proximity effect. (b) ((c))
  Real space probability amplitude $|\psi|$ across the width for the Majorana edge state 
with a smaller (larger) momentum $|k_x|$ ($k_x<0$) at one edge (i.e., $y=40000$)
(only part of the ribbon is shown). The fluctuations of $|\psi|$ at the
positions far away from the edge are due to numerical error. Here, 
$V_z=6\ $meV, $\lambda=4\ $meV, $\mu=2\ $meV and $\Delta=2\ $meV. 
}
\label{S1}
\end{figure}

We turn to the case of armchair graphene ribbon with the Hamiltonian being
\begin{widetext} 
\begin{eqnarray}
H_{\rm armchair}&=&-t\sum_{k_x}\sum_{\langle j_1,j_2
  \rangle\sigma}c^{\dagger}_{k_xj_1\sigma}c_{k_xj_2\sigma}+\sum_{k_x}\sum_{j\sigma}(\sigma
V_z-\mu)c^{\dagger}_{k_xj\sigma}c_{k_xj\sigma}+\Delta\sum_{k_x}\sum_{j}(c^{\dagger}_{k_xj\uparrow}c^{\dagger}_{-k_xj\downarrow}+{\rm
  H.c.})\nonumber\\
&&\hspace{-1.5cm}\mbox{}+i\lambda\sum_{k_x}\sum_{\langle j_1,j_2
  \rangle\sigma\sigma^{\prime}}(\sigma_x^{\sigma\sigma^{\prime}}d^y_{j_1j_2}-\sigma_y^{\sigma\sigma^{\prime}}d^x_{j_1j_2})
c^{\dagger}_{k_xj_1\sigma}c_{k_xj_2\sigma^{\prime}}
-t\sum_{k_x}\sum_{j^*_1j^*_2\sigma}[e^{i\sqrt{3}k_x}(\delta_{j^*_1,j^*_2}+\delta_{j^*_1+1,j^*_2})c^{\dagger}_{k_xj^*_1\sigma}c_{k_xj^*_2\sigma}+{\rm
H.c.}]\nonumber\\
&&\hspace{-1.5cm}\mbox{}+i\lambda\sum_{k_x}\sum_{j^*_1j^*_2\sigma\sigma^{\prime}}\{e^{i\sqrt{3}k_x}
[\delta_{j^*_1+1,j^*_2}(\frac{\sqrt{3}}{2}\sigma_x^{\sigma\sigma^{\prime}}-\frac{1}{2}\sigma_y^{\sigma\sigma^{\prime}})
-\delta_{j^*_1,j^*_2}(\frac{\sqrt{3}}{2}\sigma_x^{\sigma\sigma^{\prime}}+\frac{1}{2}\sigma_y^{\sigma\sigma^{\prime}})]
c^{\dagger}_{k_xj^*_1\sigma}c_{k_xj^*_2\sigma^{\prime}}+{\rm H.c.}\}
\end{eqnarray}
\end{widetext}
in which $j^*_1$ ($j^*_2$) represents the $j^*_1$-th ($j^*_2$-th) atom of the
first (fourth) column in the unit cell.  Note that the unit
cell of the armchair ribbon is the same as the one in Ref.~\onlinecite{brey} and
the edges lie along the $x$-direction. Similar to the case of the zigzag
graphene ribbon, we first solve the armchair ribbon with only the hopping term analytically near the Dirac
points. The eigenstates read 
\begin{eqnarray}
\Psi_{k_x}^{k_n,\varepsilon}({\bf r})&=&2Ae^{ik_xx}\sin[(|K|+k_n)y]\nonumber\\
&&\mbox{}\times\left(\begin{array}{c}
-v_f(k_x-ik_n)/\varepsilon \\
i \\
\end{array}\right),
\end{eqnarray} 
where the eigenvalues are $\varepsilon^2=v_f^2(k_x^2+k_n^2)$ with
$k_n=n\pi/L-|K|$ and $A=\frac{1}{\sqrt{8L}}$. These eigenstates construct 
complete basis functions for $H_{\rm armchair}$ with additional spin and
particle-hole degrees of freedom. By diagonalizing the Hamiltonian matrix, one
obtains the energy spectrum and eigenstates of armchair graphene ribbon as
shown in Fig.~\ref{S1}. In Fig.~\ref{S1}(a), we find that there exist eight
zero energy states, corresponding to four Majorana fermions at each edge, which
is similar to the case of zigzag ribbon. We then show the real space probability
amplitude of two Majorana edge states at the same edge with a smaller and larger momentum
$|k_x|$ ($k_x<0$) in Figs.~\ref{S1}(b) and (c), respectively. It is seen that both show
obvious decays and oscillations but the decay lengths and oscillation periods are different.

\end{appendix}

\end{document}